\documentclass[twocolumn,showpacs,preprintnumbers,amsmath,amssymb,pre,superscriptaddress]{revtex4}

\usepackage{graphicx}
\usepackage{dcolumn}
\usepackage{bm}

\begin{document}

\preprint{}

\title{Min-protein oscillations in {\em Escherichia coli} with spontaneous formation of two-stranded filaments in a three-dimensional stochastic reaction-diffusion model}

\author{Nenad Pavin}%
\affiliation{Department of Physics, Faculty of Science, University of
Zagreb, Zagreb, Croatia}
\author{Hana \surname{\v{C}ip\v{c}i\'{c} Paljetak}}
\affiliation{PLIVA Research Institute Ltd., Zagreb, Croatia}%
\author{Vladimir Krsti\'{c}}
\affiliation{Department of Physics, Faculty of Science, University of
Zagreb, Zagreb, Croatia}

\date{\today}

\begin{abstract}
We introduce a three-dimensional stochastic reaction-diffusion model to describe MinD/MinE dynamical structures in {\it Escherichia coli}. 
This model spontaneously generates  pole-to-pole oscillations of the membrane-associated MinD proteins, MinE ring, as well as filaments of the membrane-associated MinD proteins. Experimental data suggest MinD filaments are two-stranded. 
In order to model them we assume that each membrane-associated MinD protein can form up to three bonds with adjacent membrane associated MinD molecules and that MinE induced hydrolysis strongly depends on the number of bonds MinD has established.

\end{abstract}


\maketitle

A division site in the rod-shaped bacterium {\em E. coli} is determined by the location of the FtsZ-ring \cite{Lutkenhaus_02}. Two major factors known to be important for  placement of the FtsZ ring are nucleoid occlusion  and  Min-protein oscillations \cite{Yu_99}.
Nucleoid occlusion restricts possible division sites to regions void of DNA  --- near the center and  poles of the cell ---  while rapid pole-to-pole Min oscillations exclude poles as the possible division site \cite{deBoer_89,Bi_93}.

The Min system consists of three proteins: MinC, MinD, and MinE. MinD and MinE proteins generate pole-to-pole oscillations, while  MinC proteins are being recruited to the membrane by MinD and hence  follow the same oscillatory pattern \cite{Raskin_99}. Whereas MinC inhibits polymerization of FtsZ \cite{Hu_99}, pole-to-pole oscillations prevent asymmetric cell division.
MinD proteins in the ATP-bound form (MinD:ATP) attach to the membrane and presumably form two-stranded filaments  arranged into a helix \cite{Hu_02,Suefuji_02,Shih_03}. MinE proteins function as homodimers \cite{King_00}; they  attach to the membrane-associated MinD:ATP where  they induce ATP hydrolysis, releasing subsequently   MinD:ADP, MinE, and  phosphate into the cytoplasm. The released MinD:ADP cannot bind to the membrane, until a nucleotide exchange takes place \cite{Hu_02}. 

There are several analytical \cite{Meinhardt_01,Howard_01,Kruse_02,Huang_03,Kulkarni_04,Meacci_05,Drew_05}  and stochastic models \cite{Howard_03} that successfully reproduce Min oscillations. 
In this work we introduce the three-dimensional stochastic model.
Our model, in contrast to others, takes into account a finite
size of Min molecules and their spatial organization on the
membrane.


\section{The model and simulations}
\begin{figure}
\includegraphics[width=7.8cm]{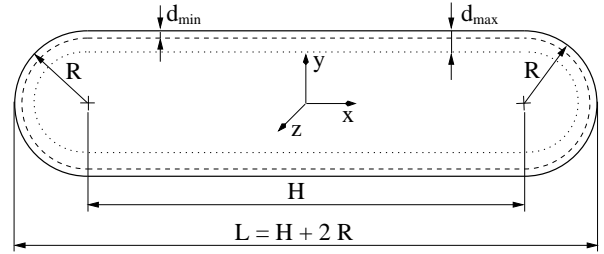}
\caption{\label{fig:bacteria} Geometric shape used to model the bacterium
{\it E. coli}: 
cylinder of length $H$ and radius $R$  with two hemispheres of radius $R$
at either end. The origin of the coordinate system ($x,y,z$) is placed
in the center of the bacterium. Two  distances from the membrane,
$d_{min}$ and $d_{max}$, are used to define the region near the membrane
($d\!<\!d_{min}$) and the region far from the membrane ($\!d>\!d_{max}$),
respectively. For details see text. All parameters shown in the figure
are  scaled equally, except the parameter $d_{min}$.} 
\end{figure}

\begin{figure}
\centering
\includegraphics[width=8.4cm]{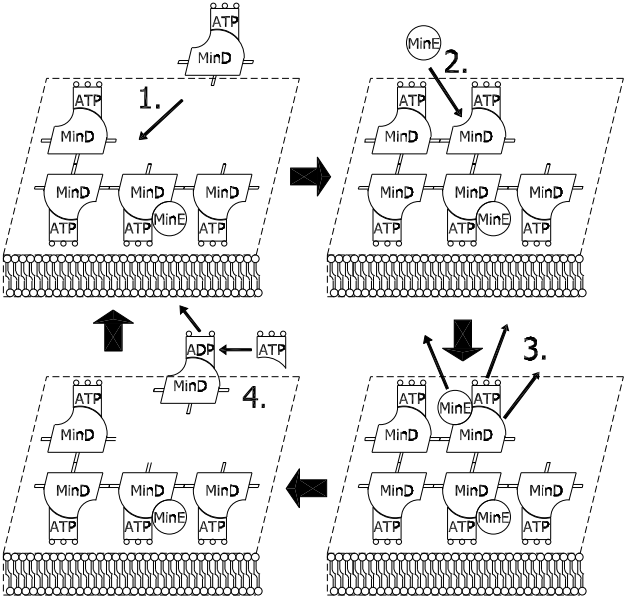}
\caption{\label{fig:four_phases}Schematic representation of four stages in MinD/MinE proteins
dynamics. (1.) MinD:ATP binds to the inner layer of the cytoplasmic membrane; (2.) MinE binds to the
membrane-associated MinD:ATP; (3.) MinE induces ATP hydrolysis; 
releasing subsequently  MinD:ADP, MinE, and phosphate into the cytoplasm, (4.)
MinD:ADP is converted back into MinD:ATP by nucleotide exchange.}
\end{figure}

The shape of the bacterium {\it E. coli} is approximated by a cylinder of
length $H$ and radius $R$  with two hemispheres of radius $R$ at either
end (two poles of the bacterium), giving the total length $L=H+2R$ 
(Fig.~\ref{fig:bacteria}). 
Experimentally observed oscillations of MinC/MinD/MinE proteins between two
poles are modeled using only MinD and MinE proteins. All interactions
included in the model take place simultaneously. 
However, for a particular MinD  these interactions occur in four successive stages
(Fig.~\ref{fig:four_phases}), which extend those proposed by Huang {\em et al.} \cite{Huang_03} by taking into account the spatial organization of Min proteins on the membrane.
%

(1.) Cytoplasmic MinD:ATP freely diffuses; When  near the membrane it tries to attach to it in two ways: (i) either independently of other MinD:ATP molecules already attached, (ii) or it tries to became a part of a double chain of MinD:ATP molecules (two-stranded filament) already formed on the membrane. 
 
(2.) MinE freely diffuses through cytoplasm. It does not attach to the
membrane nor cytoplasmic MinD. However, MinE can attach to the
membrane-associated MinD:ATP  forming a MinE-MinD:ATP complex. 

(3.) MinE protein in the membrane-associated MinE-MinD:ATP complex
stimulates detachment of the complex from the membrane by inducing ATP
hydrolysis, releasing subsequently  MinD:ADP, MinE, and  phosphate into the cytoplasm. 

(4.) The MinD:ADP complex cannot attach to the membrane until it is transformed back  into MinD:ATP by nucleotide exchange.

%

In our model each molecule is treated as a pointlike particle, except when on the membrane where the molecule  size is taken into account.

The diffusion process in  three-dimensional space is described using
Smoluchowski dynamics \cite{vonSmol_17, Andrews_04}. A molecule starts from a well-defined position
$\vec{r}$ at time $t$ and diffuses during a time $\Delta t$. Probability
density for finding a molecule at  time $t+\Delta t$ at a position
$\vec{r}+\Delta\vec{r}$ is described by 
\begin{equation}
\label{eq:gauss}
p(\vec{r}+\Delta\vec{r},t+\Delta t)=
G_{s}(\Delta x) G_{s}(\Delta y) G_{s}(\Delta z),
\end{equation}
\begin{equation}
\label{eq:gauss1}
G_{s}(\Delta x)\equiv \frac{1}{s\sqrt{2\pi}}%
\exp\left(-\frac{(\Delta x)^2}{2s^2}\right),
\end{equation}
\begin{equation}
\label{eq:gauss2}
s\equiv\sqrt{2D\Delta t},
\end{equation}
where $G_{s}(\Delta x)$ is a normalized Gaussian distribution with deviation
$s$ (diffusion length), and diffusion coefficient $D$.

The position of the molecule at time $t+\Delta t$ is obtained by adding random displacement to the current position where 
distribution of random displacements  obeys  (\ref{eq:gauss}).
Generally, time steps $\Delta t$ do not have to be kept constant.
In our simulation we use adaptive $\Delta t$ in order to focus computational effort on important time segments.  Because diffusion and unimolecular reactions are the only processes that take place in the region far away from the membrane, one can use longer time steps in that region than in the region near the membrane where, in addition, bimolecular processes occur. These two regions are defined using two free model parameters $d_{min}$ and $d_{max}$ --- characteristic distances from the membrane (Fig.~\ref{fig:bacteria}).
In the region far away from the membrane ($d\!>\!d_{max}$)
time step used is significantly longer
than time step used in the region near the membrane ($d\!<\!d_{min}$). 
In the transitional region
($d_{min}\!<\!d\!<\!d_{max}$) time step is gradually decreased when
approaching the membrane, to avoid that molecules entering the region
near the membrane diffuse too far, avoiding on their path 
bimolecular reactions. For the same reason, time step $\Delta t$ in the region
near the membrane has to be chosen such that condition
$s\equiv\sqrt{2D\Delta t}\ll d_{min}$ is satisfied. 

In our model, parameter $d_{min}$ is also  used as the reaction radius parameter for all bimolecular reactions.
Hence, the cytoplasmic MinD:ATP molecule can  attach to the membrane only
when it is in the region near the membrane.
Probability for this reaction is given by the simple intuitive formula:
\begin{equation}
\label{eq:D}
p_{D}=\sigma_{D}\frac{\Delta t}{d_{min}}.
\end{equation}
The probability is proportional to time step $\Delta t$ --- the
longer you wait it is more probable for a reaction to take place --- and inversely proportional to $d_{min}$ to ensure that the number of reactions taking place depends only on the reaction rate parameter $\sigma_{D}$ and not on the model parameter $d_{min}$ 
used to define the near membrane region. 
If the reaction occurs, the molecule attaches to the membrane with random orientation. However, our model forbids this reaction to take
place if the position where the molecule should bind is already occupied 
by another MinD:ATP.

Additionally, cytoplasmic MinD:ATP can react with MinD:ATP molecules already attached to the membrane. Experimental data suggest that MinD:ATP attached to the membrane polymerizes  into two-stranded filaments \cite{Hu_02,Suefuji_02,Shih_03}. 
Lacking experimental data on the interaction between membrane-associated MinD:ATP molecules, we  utilize the simplest assumption in which each MinD:ATP molecule can form up to three bonds with  adjacent MinD:ATP molecules (Fig.~\ref{fig:four_phases}).
The probability for cytoplasmic MinD:ATP to  occupy any  free attachment site that is within reaction radius ($r<d_{min}$) depends on the reaction rate $\sigma_{Dd}$:
\begin{equation}
\label{eq:Dd}
p_{Dd}=\sigma_{Dd}\frac{\Delta t}{V}; \qquad V=\frac{2\pi}{3}d_{min}^{3}.
\end{equation} 

An attachment of MinE to the membrane-associated MinD:ATP complex can take place if molecules are within the interaction radius ($r<d_{min}$) and there is no MinE molecule already attached. The probability for this reaction  is
\begin{equation}
\label{eq:E}
p_{E}=\sigma_{E}\frac{\Delta t}{V}; \qquad V=\frac{2\pi}{3}d_{min}^{3}.
\end{equation}

MinE protein in the membrane-associated MinE-MinD:ATP complex stimulates
detachment of the complex from the membrane by inducing ATP hydrolysis.
The probability for this reaction might depend on the number of bonds particular MinD:ATP has formed with its MinD:ATP neighbors, and we assume that the number of bonds established decreases the reaction probability.
Let $\sigma_{de}^{(i)}$, $i=0,1,2,3$ stand for the detachment reaction rate when the MinD molecule has $i$ bonds established. Hence,
\begin{equation}
\label{eq:de}
\sigma_{de}^{(0)} > \sigma_{de}^{(1)} > \sigma_{de}^{(2)} > 
\sigma_{de}^{(3)},
\end{equation}  
and the probability for this reaction is
\begin{equation}
\label{eq:hydro}
p_{de}^{(i)}= 1-\exp\left({-\sigma_{de}^{(i)}\Delta t}\right).
\end{equation}

The transformation of MinD:ADP into MinD:ATP by 
nucleotide exchange  is treated as unimolecular
reaction with reaction rate $\sigma_{D}^{ADP \rightarrow ATP}$. Hence, the
probability for this reaction during time interval $\Delta t$ is
\begin{equation}
\label{eq:ADP2ATP}
p_{D}^{ADP \rightarrow ATP}= 1-\exp\left({-\sigma_{D}^{ADP \rightarrow ATP}\Delta t}\right).
\end{equation}
\begin{figure}
\includegraphics[width=6.0cm,angle=-90]{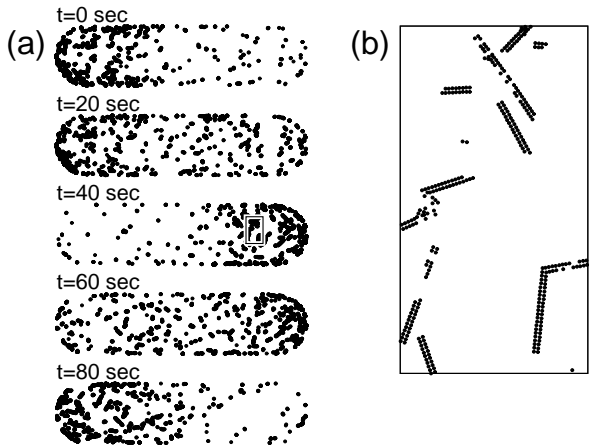}
\caption{\label{fig:ortgoProjA}(a) Orthogonal projection of the
membrane-associated MinD proteins 
onto a plane parallel to the line connecting two poles of the bacterium.
Each projected protein is represented with a dot. Five time frames are
shown. They refer to times 
$0, \frac{1}{4}T, \frac{2}{4}T, \frac{3}{4}T, T$; where
$T\approx 80~\mbox{sec}$ is
the period of oscillation obtained with parameters specified in the text.
(b) A portion of (a) is enlarged to clearly show two-stranded  filaments.} 
\end{figure}
%
%
\begin{figure}
\includegraphics[width=7.5cm]{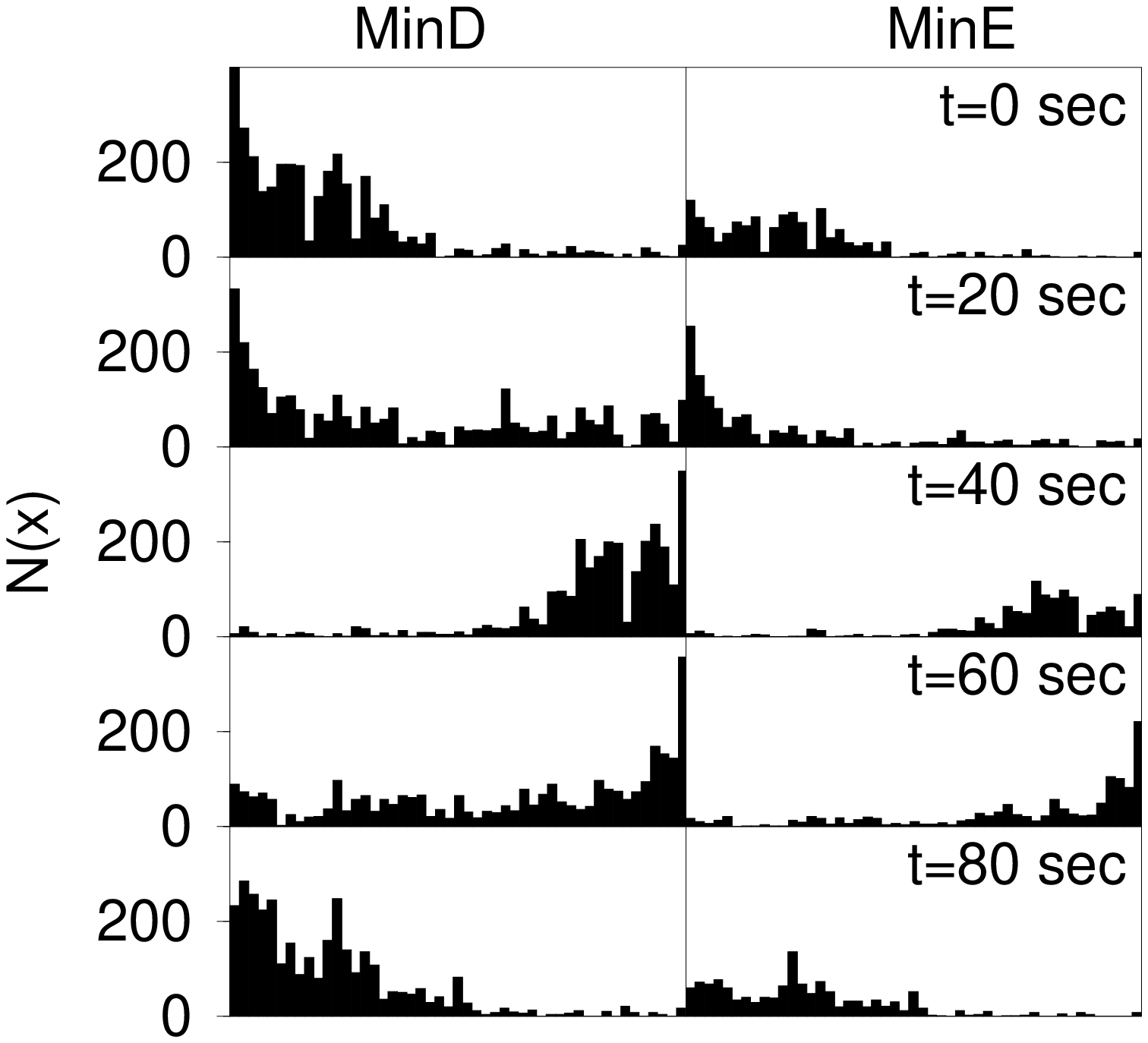}
\includegraphics[width=7.5cm]{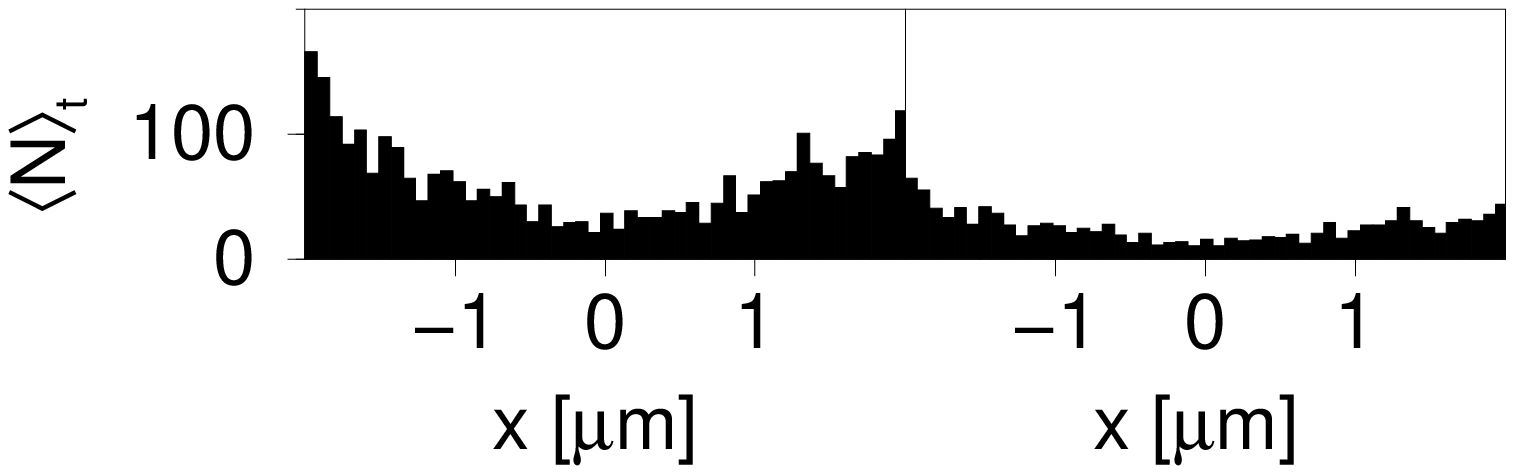}
\caption{\label{fig:histograms}Histogram for the number of the membrane-associated MinD (MinE)
proteins versus $x$-coordinate, for the same time frames as in
Fig.~\ref{fig:ortgoProjA}. Size of the  bin used is $0.08~\mu \mbox{m}$. Time average of the number of the membrane-associated MinD (MinE) proteins  over three periods is given in the last row.}
\end{figure}
%

\section{Results and Discussion}
In our numerical simulations we have fixed parameters related to the
geometry of the cell (Fig.~\ref{fig:bacteria}) to $R\!=\!0.5\,\mu \mbox{m}$
(one-half cell width) and $L=4\,\mu \mbox{m}$ (cell length). 
Diffusion constants for cytoplasmic MinD and MinE proteins that are used,  are in good agreement with measured values for {\em E. coli} proteins of similar size \cite{Elowitz_99}:
\begin{equation}
D_{D}=D_{E}=2.5\,\mu \mbox{m}^2/\mbox{sec}.
\end{equation}
Reaction rate parameters are chosen such that experimentally observed oscillations are reproduced:
\begin{eqnarray}
~~\sigma_{D}=0.01\,\mu \mbox{m}/\mbox{sec}, \;\;
\sigma_{Dd}=0.003\,\mu \mbox{m}^{3}/\mbox{sec}, \nonumber \\
\sigma_{E}=0.02\,\mu \mbox{m}^{3}/sec, \;\;
\sigma_{D}^{ADP \rightarrow ATP}= 1/\mbox{sec}.
\end{eqnarray}
These parameters are similar to the parameters that Huang {\em et al.} \cite{Huang_03} used in their analytical model.
However, they  use only one hydrolysis rate
parameter, while in our model there are four  ---
$\sigma_{de}^{(i)}$, $i=0,1,2,3$ --- which obey (\ref{eq:de}) with the
ratio:
\begin{eqnarray}
\label{eq:ratio}  
\sigma_{de}^{(0)}:\sigma_{de}^{(1)}:\sigma_{de}^{(2)}: 
\sigma_{de}^{(3)} = 540:135:45:1 \nonumber \\ 
\sigma_{de}^{(0)}=7.2~\mbox{sec}^{-1}.
\end{eqnarray} 
Other ratios have been tested also. It is found that it is essential to
take $\sigma_{de}^{(3)}$ significantly smaller than $\sigma_{de}^{(2)}$
in order to generate oscillations, whose period primarily  (and strongly) depends on the parameter $\sigma_{de}^{(0)}$.  In the case when all
$\sigma_{de}^{(i)}$ were taken to be identical, the oscillations could not be produced even if other parameters of the model were varied substantially.

In our simulation we use 4000 MinD molecules and 1400 MinE homodimers, reflecting the {\em in vivo} situation \cite{Shih_02}.
The two-stranded filament width and the MinD  monomer length are fixed to 6 and  5~nm, respectively \cite{Suefuji_02}. The model is evolved in time with time step $\Delta t\!=\!4\cdot 10^{-5}$~sec for processes far from the membrane ($d_{max}\!=\!0.1$~$\mu$m) and  $\Delta t\!=\!8\cdot 10^{-7}$~sec for processes near the membrane ($d_{min}\!=\!0.01$~$\mu$m). In the transitional region time steps are gradually decreased when approaching the membrane. 
We have tested the simulation by significantly varying parameters $d_{min}$ and $\Delta t$ and the same results were obtained.

With these parameters we have reproduced pole-to-pole MinD/MinE oscillations (Figs.~\ref{fig:ortgoProjA}~and~\ref{fig:histograms}) with the period $T\approx\!80$~sec, which is compatible with experimentally observed range ($30-120$~sec) \cite{Raskin_99}. 
Initially, all MinE and MinD  are placed in the center of the bacterium. Other initial distributions were tried (e.g., uniform distribution), and the same type of oscillations always appeared after the transient period lasting up to one oscillation cycle.
  
Distributions of the membrane-associated MinD/MinE proteins do not oscillate in phase ---  MinE distribution lags after MinD distribution (Fig.~\ref{fig:histograms}). This phenomenon has been seen in experiments, and it was described as an oscillating MinE ring \cite{Shih_02}.  
However, when time-averaged both distributions  have a minimum in the middle of the cell (last row in the Fig.~\ref{fig:histograms}) which reflects distribution necessary for proper cell division.
This minimum is experimentally observed only in the case of the MinD protein
oscillation \cite{Meacci_05}. For the time-averaged MinE distribution there are only model
predictions and they disagree on this point; e.g., there are models which predict,
in contrast to our model prediction, that the time-averaged MinE distribution
has a maximum in the middle of the cell \cite{Howard_01,Howard_03}. 
\begin{figure}
\includegraphics[width=3.0cm,angle=-90]{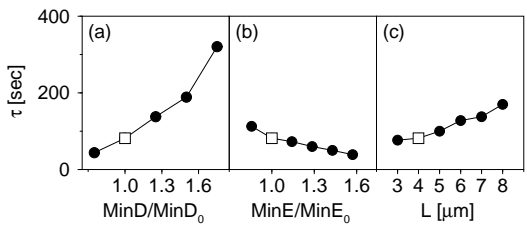}
\caption{\label{fig:overex}
Dependence of the oscillation period on
the  total number of MinD and MinE molecules, and the cell length.
Points marked by $\square$ are obtained with parameters used to generate Fig.~\ref{fig:ortgoProjA}.
In (a) and (b) parameters are varied one at a time while keeping the others fixed. In (c), instead of keeping the number of MinD and MinE fixed, their concentrations are fixed at values used in Fig.~\ref{fig:ortgoProjA} while cell length is varied.} 
\end{figure}
\begin{figure}
\includegraphics[width=5.0cm,angle=-90]{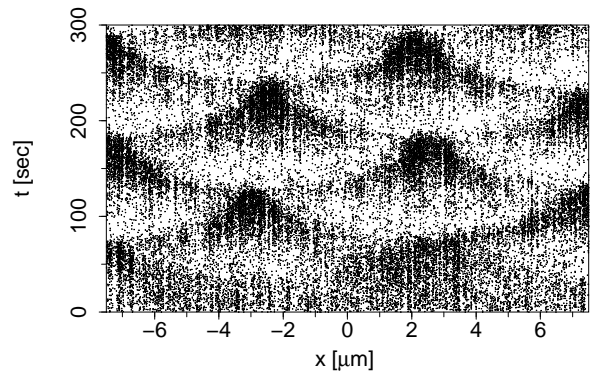}
\caption{\label{fig:filament}
Space-time plot of the membrane-associated MinD proteins in the filamentous cell ($L\!=\!15~\mu$m ). Each MinD is represented with a dot. For the sake of clearness, we have reduced the number of dots by a factor of 100.
Concentrations of MinD and MinE, plus all the remaining model parameters are fixed at values used to obtain results in  Fig.~\ref{fig:ortgoProjA}.}
\end{figure}
\begin{figure}
\includegraphics[width=1.8cm,angle=-90]{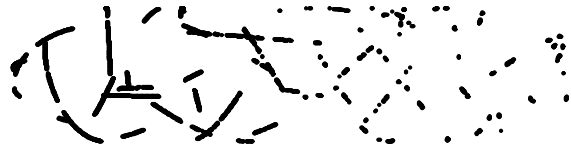}
\caption{\label{fig:longerFil}
Membrane-associated MinD proteins oscillations; only a single time frame is
given. Parameter values are the same as those used in
Fig.~\ref{fig:ortgoProjA}, except $\sigma_{Dd}=0.02\,\mu
\mbox{m}^{3}/\mbox{sec}$, 
$\sigma_{D}=0.0005\,\mu \mbox{m}/\mbox{sec}$, and
$\sigma_{de}^{(0)}=11.0~\mbox{sec}^{-1}$ [parameters $\sigma_{de}^{(1)}$,
$\sigma_{de}^{(2)}$, $\sigma_{de}^{(3)}$ were modified to obey
(\ref{eq:ratio})]. MinD filaments are longer then those obtained in 
Fig.~\ref{fig:ortgoProjA}.} 
\end{figure}

To confirm its robustness, the model was additionally tested for a variety of experimentally observed phenomena. Fu {\em et al.} \cite{Fu_01} have found experimentally that the oscillation period increases with the cell length. The overexpression experiments reveal that the oscillation period increases with the amount of MinD, and decreases with the amount of MinE \cite{Raskin_99}. All these phenomena are reproduced with our model (Fig.~\ref{fig:overex}). Consistent with experiments \cite{Raskin_99}, in the case of filamentous cells
the zebra-striped oscillation pattern is obtained spontaneously (Fig.~\ref{fig:filament}), starting with uniform distributions of MinD and MinE.

In our model, MinD proteins attached to the membrane  predominately form  two-stranded filaments [Fig.~\ref{fig:ortgoProjA}(b)].
This is achieved by imposing the ratio (\ref{eq:ratio}) to the parameters  $\sigma_{de}^{(i)}$ which are responsible for dynamics of both formation and decomposition of two-stranded filaments. The imposed  ratio strongly favors two-stranded configurations over a group of single molecules ---  the probability for a group of single molecules to be detached from the membrane is considerably greater than that for the same group of molecules, but in the form of the two-stranded filament.

The filament appears as an alive object. It is degraded and
rebuilt constantly. When it grows in size the building process
dominates over the degrading process. Both processes preferentially
take place at the filament's end. MinD molecules
located at the end of the filament can form one or two bonds
with its neighbors, while other MinD molecules have probably
established three bonds. Because of (\ref{eq:ratio}) it is more probable
for MinE to detach MinD molecules locate the filament's
end. If the building process dominates over the
degrading process, detached MinD molecules will probably
be replaced with cytoplasmic MinD:ATP molecules.
       
However, as the concentration of MinD molecules attached
to the membrane reaches its peak, the concentration
of cytoplasmic MinD:ATP goes to its minimum. At that time
the degrading process dominates over the building process.
Cytoplasmic MinE molecules continue to attach to the MinD
molecules of the two-stranded filament. MinD released into
cytoplasm by MinE is in the form of the MinD:ADP complex
and cannot bind to the membrane. However, MinE released
into the cytoplasm by the same process attach to free
attachment sites on the filament, thus speeding up its decomposition.

The average length of filaments obtained with our model depends on the
model parameters, particularly on  $\sigma_{Dd}$ and  $\sigma_{D}$. If we increase
the parameter $\sigma_{Dd}$ and/or decrease the parameter $\sigma_{D}$, the 
probability for attaching MinD to the filament already formed on the
membrane (regulated by $\sigma_{Dd}$) will increase with respect to the
probability for starting a new filament formation (regulated by
$\sigma_{D}$). Hence, the average length of filaments is increased
(Fig~\ref{fig:longerFil}).
In  order to keep the period of oscillation similar to the period for the case
shown in Fig.~\ref{fig:ortgoProjA}, hydrolysis rate parameters were
modified: $\sigma_{de}^{(0)}=11.0~\mbox{sec}^{-1}$, while the same ratio
(\ref{eq:ratio}) was obeyed.  

In conclusion, we have introduced 3D stochastic reaction-diffusion model to 
describe  MinD/MinE dynamical structures in {\it E. coli}. In particular, our
model spontaneously generates two-stranded filaments using a few simple physical
assumptions.

\section{Acknowledgments}

We thank Ivana \v Sari\' c for a technical support, and Goran Mitrovi\' c for
carefully reading our manuscript.


\end{document}